*Research Article*

# High Order QCD Predictions for Inclusive Production of *W* Bosons in *pp* Collisions at $\sqrt{s} = 13$ TeV

**Hasan Ogul,[1,2] Kamuran Dilsiz,[1] Emrah Tiras,[1] Ping Tan,[3]**
**Yasar Onel,[1] and Jane Nachtman[1]**

[1]*Department of Physics and Astronomy, University of Iowa, Iowa City, Iowa 52242, USA*
[2]*Department of Physics, Sinop University, Korucuk, 57000 Sinop, Turkey*
[3]*Department of Physics and Astronomy, University of Rochester, Rochester, NY 14627, USA*

Correspondence should be addressed to Hasan Ogul; hsnogul@gmail.com

Predictions of fiducial cross sections, differential cross sections, and lepton charge asymmetry are presented for the production of $W^{\pm}$ bosons with leptonic decay up to next-to-next-to-leading order (NNLO) in perturbative QCD. Differential cross sections of $W^{\pm}$ bosons and $W$ boson lepton charge asymmetry are computed as a function of lepton pseudorapidity for a defined fiducial region in $pp$ collisions at $\sqrt{s} = 13$ TeV. Numerical results of fiducial $W^{\pm}$ cross section predictions are presented with the latest modern PDF models at next-to-leading order (NLO) and NNLO. It is found that the CT14 and NNPDF 3.0 predictions with NNLO QCD corrections are about 4% higher than the NLO CT14 and NNPDF 3.0 predictions while MMHT 2014 predictions with NLO QCD corrections are smaller than its NNLO QCD predictions by approximately 6%. In addition, the NNLO QCD corrections reduce the scale variation uncertainty on the cross section by a factor of 3.5. The prediction of central values and considered uncertainties are obtained using FEWZ 3.1 program.

## 1. Introduction

Inclusive production of $W$ bosons in proton-proton ($pp$) collisions is an important ingredient in a variety of precision studies of Standard Model (SM) parameters and derived quantities as well as a key element in Large Hadron Collider (LHC) detector calibration. The leading order processes for inclusive $W$ boson production in $pp$ collisions are mainly from $u\bar{d} \rightarrow W^{+}$ and $d\bar{u} \rightarrow W^{-}$. Due to the presence of two $u$ valance and one $d$ valance quarks in the proton, $W^{+}$ bosons are produced more often than $W^{-}$ in $pp$ collisions. This difference leads to an asymmetry between $W^{+}$ and $W^{-}$, which constrains the ratio of $u$ and $d$ quarks. $W$ boson production asymmetry is defined to be

$$A_W(y) = \frac{\sigma(W^{+}) - \sigma(W^{-})}{\sigma(W^{+}) + \sigma(W^{-})}. \quad (1)$$

Due to cancellation of both experimental and theoretical uncertainties, a high precision measurement of this asymmetry as a function of boson rapidity ($y$) can be used to improve the current knowledge of parton distribution functions (PDFs). However, it is difficult to measure the production asymmetry because of the energy carried away by neutrinos in leptonic $W$ boson decays. A quantity more directly accessible experimentally is the lepton charge asymmetry, defined as

$$A(\eta)$$
$$= \frac{(d\sigma/d\eta)(W^{+} \rightarrow l^{+}\nu) - (d\sigma/d\eta)(W^{-} \rightarrow l^{-}\nu)}{(d\sigma/d\eta)(W^{+} \rightarrow l^{+}\nu) + (d\sigma/d\eta)(W^{-} \rightarrow l^{-}\nu)}, \quad (2)$$

where $d\sigma/d\eta$ is the differential cross section for $W$ boson production and subsequent leptonic decay and $\eta$ is the charged lepton pseudorapidity which encompasses the beam angle.

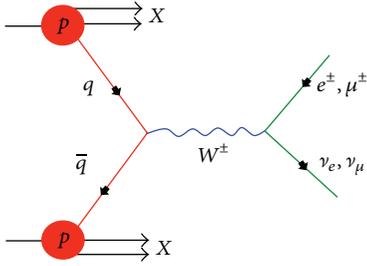

FIGURE 1: Leading order $W$ production diagram in $pp$ collisions.

Both $W$ boson lepton charge asymmetry and $W$ boson production charge asymmetry were previously studied in $p\bar{p}$ collisions by the CDF and D0 collaborations [1–5]. At the LHC, CMS, ATLAS, and LHCb experiments performed $W$ boson lepton charge asymmetry using the data collected in 2010 [6–8]. The CMS experiment has further improved the measurement precision by using data collected at 7 TeV (in 2011) and at 8 TeV (in 2012) center of mass energies ($\sqrt{s}$). The number of $W \rightarrow \mu\nu$ events in the 8 TeV measurement [9] (more than 100 million) is 5 orders of magnitude larger than the events in the 7 TeV measurement [10]. In addition, the average value of the Bjorken scaling variable [11] for the interacting partons in $W^{\pm}$ boson production at $\sqrt{s} = 8$ TeV is lower than at $\sqrt{s} = 7$ TeV, which is expected to result in a lower $W$ boson production charge asymmetry. The average value of this parameter is smaller for $\sqrt{s} = 13$ TeV data collected in 2015 at LHC, and this let us study the PDFs for lower $x$ values. The measurement precision could be further improved using 13 TeV data, and this may help to improve the determination of the PDFs.

$W$ processes shown in Figure 1 are well understood and can be used to test precision predictions based on quantum chromodynamics (QCD). In addition to the lepton charge asymmetry predictions, inclusive and differential cross section measurements of $W^+$ and $W^-$ bosons are also considered in this paper. The cross section measurements of leptonic channels can provide fundamental tests of perturbative QCD since they provide clean signatures with relatively low background. Differential cross section values are also essential inputs for the calculation of the lepton charge asymmetry. The scale, PDFs, and strong coupling constant ($\alpha_s$) uncertainties are included in the measurements since they have become a limiting factor for the precision of many inclusive and differential cross section calculations. The changes in the uncertainties by center of mass energy increases are also considered in the discussions of NLO and NNLO QCD correction effects on the uncertainties.

This paper presents the theoretical predictions of the fiducial cross sections, differential cross sections, and the lepton charge asymmetry obtained using the FEWZ 3.1 [12] interfaced with the most recent PDF sets (CT14 [13], NNPDF 3.0 [14], and MMHT 2014 [15]) to understand the impact of the QCD corrections at 13 TeV center of mass energy. No electroweak corrections are included in these calculations.

DYNNLO [16] parton level Monte Carlo program, which computes the cross sections for vector boson production in $pp$ and $p\bar{p}$ collisions up to NNLO in QCD perturbation theory, is used to validate the FEWZ results.

## 2. Predictions of Fiducial Cross Sections and Lepton Charge Asymmetry

In practice, the observed cross sections and lepton charge asymmetry ($A(\eta)$) should be compared to the cross section and asymmetry predictions calculated to the highest order that is available. Currently, the highest order in perturbative QCD is NNLO. The necessity of NNLO cross section predictions is demonstrated in Figure 2. CMS [17–19] and ATLAS [20, 21] results are compared to LO, NLO, and NNLO predictions; a good description of experimental results with NNLO QCD predictions is observed although CMS and ATLAS results at 13 TeV are preliminary. LO, NLO, and NNLO PDFs are used for the LO, NLO, and NNLO QCD predictions, respectively. The colorful bands describe only PDF uncertainty. The error bars on the experimental results are only systematic uncertainties reported by the collaborations. The results show that NNLO corrections increase the predicted cross section values.

A fiducial region is further defined for calculations of $\sigma_{W^{\pm}}^{\mathrm{Fid}}$, $(d\sigma/d\eta)(W^{\pm} \rightarrow l^{\pm}\nu)$, and $A(\eta)$. The lepton is required to have $|\eta^l| < 2.4$ and $p_T^l > 25$ GeV. All presented results are for born level (pre-QED FSR) leptons. CT14, NNPDF 3.0, and MMHT 2014 PDF models are used with the corresponding value of $\alpha_s(M_Z) = 0.118$ at NLO and NNLO, and $M_W = 80.403$ GeV. The dependence of $W^+$ and $W^-$ total cross sections and the lepton charge asymmetry on $\alpha_s(M_Z)$ is illustrated at NLO and NNLO in Figure 3. It is found that the predicted cross sections increase as $\alpha_s(M_Z)$ increases. On the other hand, the asymmetry decreases as $\alpha_s(M_Z)$ increases.

Another objective of this paper is the calculation of the theoretical uncertainties on the cross sections and the lepton charge asymmetry. The PDF, scale (factorization and renormalization scales), and $\alpha_s(M_Z)$ uncertainties are considered. The PDF uncertainty is estimated following closely the prescription of the PDF4LHC working group [22, 23]. The PDF uncertainties for the CT14 and MMHT 2014 are calculated with their eigenvector sets using asymmetric master equations, and the standard deviations over its 100 replicas are evaluated for the NNPDF PDF uncertainties.

The default $\alpha_s$ value is chosen as 0.118 in the measurements. To calculate $\alpha_s$ uncertainty, the cross sections and asymmetry predictions are computed with 0.117, 0.118, and 0.119 $\alpha_s$ values. $\alpha_s$ value is consistently varied in the QCD perturbative calculations as well as in the PDF. Then, $\alpha_s$ uncertainties on the observables are calculated as follows:

$$\Delta\alpha_s = \max\left(X_{\alpha_s = 0.118} - X_{\alpha_s = 0.117}, X_{\alpha_s = 0.119} - X_{\alpha_s = 0.118}\right), \quad (3)$$

where $X$ can be cross section, differential cross section, or asymmetry.

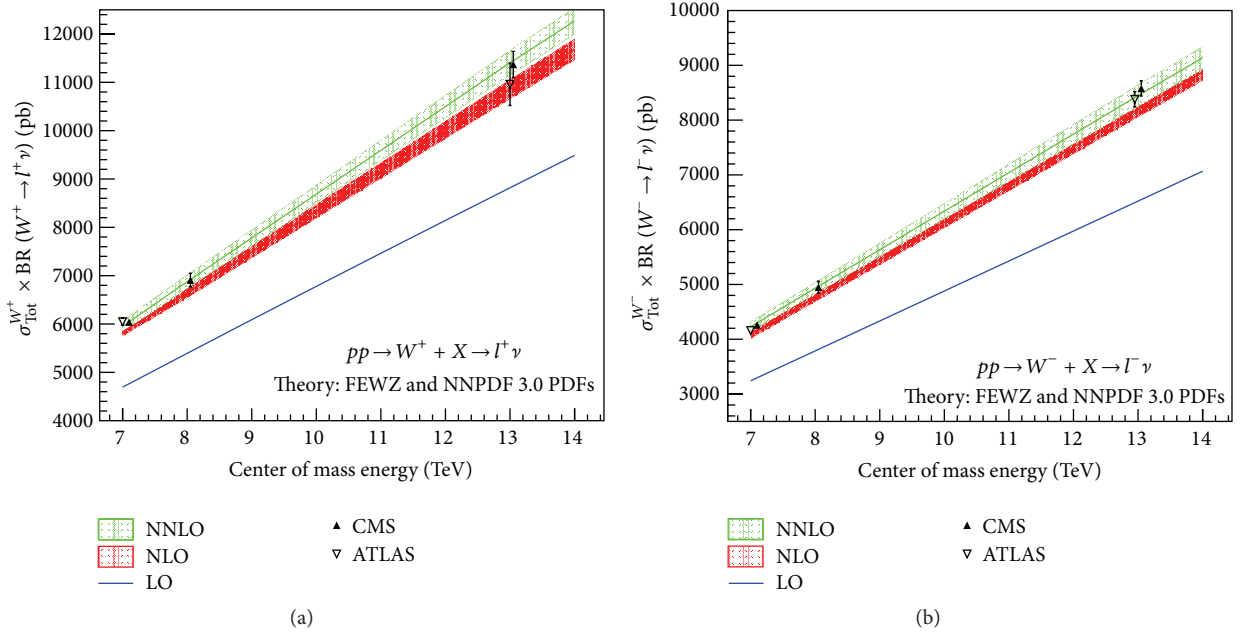

Figure 2: Experimental results of total $W$ production cross section are compared with theoretical predictions. 13 TeV CMS and ATLAS collaboration results are preliminary.

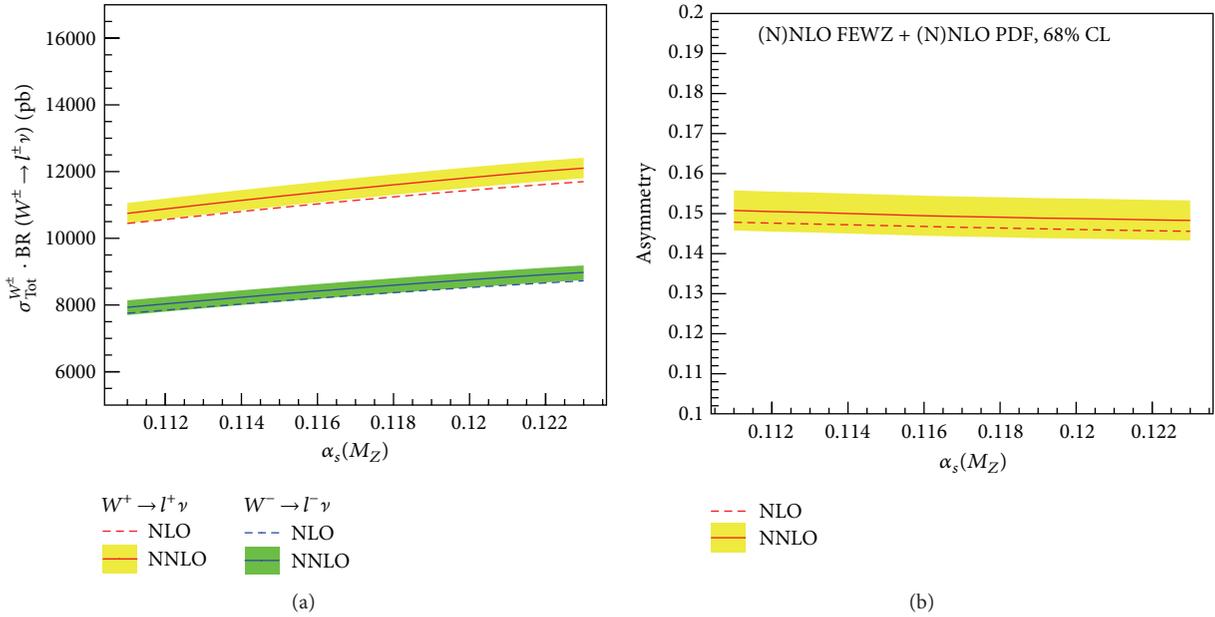

Figure 3: $W^\pm (l^\pm \nu)$ total cross sections and the lepton charge asymmetry predictions as a function of $\alpha_s$ with NLO and NNLO QCD corrections at $\sqrt{s} = 13$ TeV. The predictions are computed using CT14 NLO and NNLO PDF models.

In the measurements, the factorization and renormalization scales are chosen to be $\mu \equiv \mu_F = \mu_R = M_W$, with a theoretical uncertainty estimated by varying the scale choice by factor in the range $[1/2, 2]$. The scale variation is done simultaneously and coherently for renormalization and factorization scales. The observables are recomputed with the new factorization and renormalization scales. Then, scale uncertainties for each observable are calculated as follows:

$$\Delta\text{Scale} = \max\left(X_{\mu = M_W} - X_{\mu = M_W/2}, X_{\mu = 2M_W} - X_{\mu = M_W}\right), \quad (4)$$

where $X$ can be cross section, differential cross section, or asymmetry.

$\alpha_s$ and scale uncertainties are reported as the maximum between the positive and negative variations although they are asymmetric. The differences between positive and negative variations are slightly different. Each individual uncertainty is added in quadrature to calculate the total uncertainty. Total uncertainty means the sum of the PDF, $\alpha_s$, and scale uncertainties in quadrature throughout this paper unless otherwise specified. $\sigma^{\text{Fid}}_{W^\pm}$ and its uncertainties are presented at different perturbative orders with different PDF models in Table 1. The results show that the inclusive NNLO corrections result in approximately 4% increases on the CT14 and NNPDF 3.0 and approximately 6% increase on the MMHT 2014 NLO cross sections. In addition, it is found that the NNLO QCD corrections reduce the scale variation uncertainty by a factor of 3.5. There is no significant change on the PDF and $\alpha_s$ uncertainties.

There are many differences in the PDF analyses between CT14, NNPDF 3.0, and MMHT 2014 groups such as input data, treatments of heavy quarks, values of heavy quark masses, and ways of parameterizing PDFs, but in general it is expected that there is good agreement among them. Therefore, correlation ellipses between $W^+$ and $W^-$ fiducial cross sections are drawn for NLO and NNLO QCD predictions. Figure 4 shows the predicted fiducial cross sections times leptonic branching ratios, $\sigma_{W^+}$ versus $\sigma_{W^-}$. The ellipses illustrate the 68% CL coverage for total uncertainties. The ellipses for different PDF models show that NNLO (NLO) QCD predictions are consistent with other NNLO (NLO) PDF model predictions. The predictions with NLO QCD corrections have larger ellipses due to the larger total uncertainties.

Predictions of the fiducial differential cross section and the lepton charge asymmetry can give more information on the impact of the NNLO QCD corrections. In the fiducial region, the cross sections are evaluated differentially and the lepton charge asymmetry is obtained as a function of lepton pseudorapidity at different perturbative orders. The lepton $\eta$ binning is chosen as follows:

$$[0, 0.2, 0.4, 0.6, 0.8, 1.0, 1.2, 1.4, 1.6, 1.85, 2.1, 2.4]. \quad (5)$$

The PDF, scale, and $\alpha_s$ uncertainties are calculated for each pseudorapidity bin. The total uncertainty is determined by summing each uncertainty in quadrature. The predicted differential cross section values with their total theoretical uncertainties are presented in Table 2. The statistical precision of the predictions is 0.04%. The impact of the NNLO QCD correction as a function of lepton pseudorapidity is further investigated by examining the $k$-Factor distributions. The $k$-Factor is defined as

$$k\text{-Factor} = \frac{\left(d\sigma^{\text{Fid}}_{W^\pm}/d\eta\right)_{\text{NNLO}}}{\left(d\sigma^{\text{Fid}}_{W^\pm}/d\eta\right)_{\text{NLO}}}. \quad (6)$$

Figure 5 presents the $k$-Factor values for each $\eta$ bin. The plots show that NNLO CT14 and NNPDF 3.0 predictions are approximately 4% higher than their NLO predictions and MMHT 2014 differential cross section values are amounted to about 6% increases on the NLO fiducial cross section predictions. There is no significant $\eta$ dependence on the changes. The impact of NNLO correction on the differential cross sections is similar to what was observed on $\sigma^{\text{Fid}}$ predictions discussed earlier.

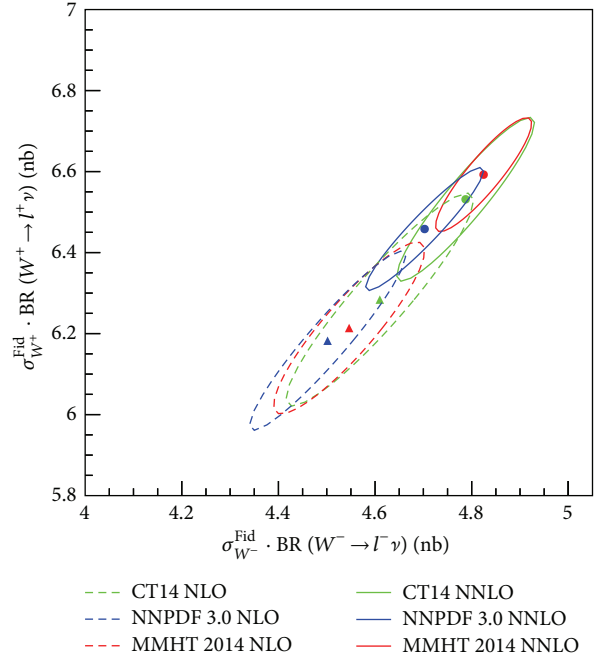

Figure 4: Predicted fiducial cross sections times leptonic branching ratios at $\sqrt{s} = 13$ TeV, $\sigma_{W^+}$ versus $\sigma_{W^-}$. The ellipses illustrate the total uncertainties.

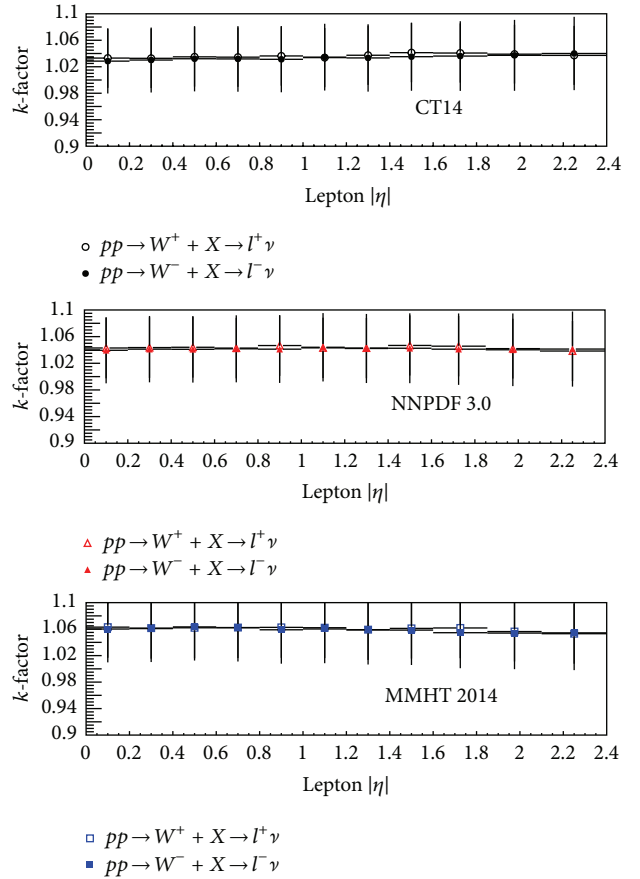

Figure 5: $k$-Factor distributions for $W^+$ and $W^-$ at $\sqrt{s} = 13$ TeV.

Table 1: Predicted fiducial cross section numbers with NLO and NNLO QCD corrections at $\sqrt{s}$ = 13 TeV. All numbers are in units of pb.

| | NLO | | | NNLO | | |
|---|---|---|---|---|---|---|
| | CT14 | NNPDF 3.0 | MMHT 2014 | CT14 | NNPDF 3.0 | MMHT 2014 |
| | | | Values for $W^+ \to l^+\nu$ (pb) | | | |
| $\sigma^{\text{Fid}}$ | 6283.9 | 6182.4 | 6213.8 | 6531.5 | 6458.2 | 6592.4 |
| PDF uncertainty | $^{+158.6}_{-186.7}$ | $\pm 112.7$ | $^{+86.9}_{-88.8}$ | $^{+174.2}_{-184.8}$ | $\pm 126.7$ | $^{+108.8}_{-86.5}$ |
| Scale uncertainty | $\pm 174.5$ | $\pm 178.2$ | $\pm 179.1$ | $\pm 49.9$ | $\pm 47.7$ | $\pm 51.6$ |
| $\alpha_s$ uncertainty | $\pm 62.1$ | $\pm 67.0$ | $\pm 71.2$ | $\pm 65.2$ | $\pm 68.9$ | $\pm 72.7$ |
| Total uncertainty | $^{+243.9}_{-262.9}$ | $\pm 221.2$ | $^{+211.4}_{-212.2}$ | $^{+192.6}_{-202.3}$ | $\pm 151.9$ | $^{+140.6}_{-124.2}$ |
| | | | Values for $W^- \to l^-\nu$ (pb) | | | |
| $\sigma^{\text{Fid}}$ | 4609.8 | 4501.7 | 4546.3 | 4787.4 | 4702.5 | 4824.6 |
| PDF uncertainty | $^{+108.8}_{-143.7}$ | $\pm 82.2$ | $^{+59.4}_{-68.7}$ | $^{+117.1}_{-142.6}$ | $\pm 104.8$ | $^{+73.1}_{-69.5}$ |
| Scale uncertainty | $\pm 122.5$ | $\pm 128.8$ | $\pm 130.4$ | $\pm 30.8$ | $\pm 34.0$ | $\pm 39.3$ |
| $\alpha_s$ uncertainty | $\pm 45.9$ | $\pm 52.5$ | $\pm 52.4$ | $\pm 47.8$ | $\pm 52.5$ | $\pm 54.8$ |
| Total uncertainty | $^{+170.2}_{-194.4}$ | $\pm 161.6$ | $^{+152.6}_{-156.5}$ | $^{+130.2}_{-153.5}$ | $\pm 122.0$ | $^{+99.5}_{-96.9}$ |

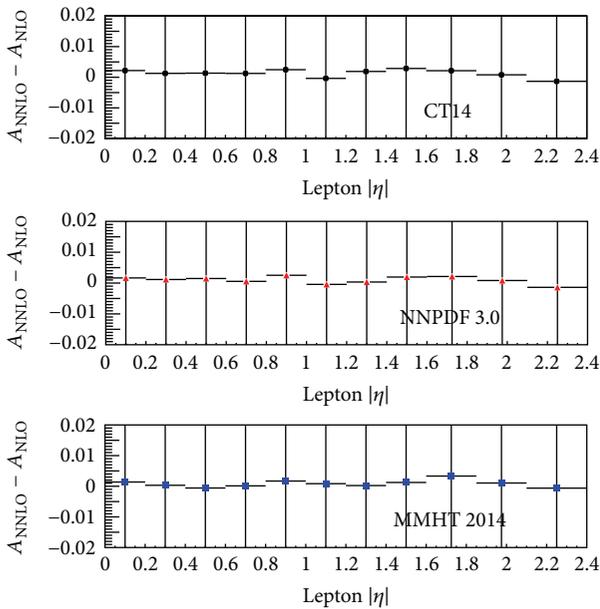

Figure 6: Impact of the NNLO correction to the asymmetry with NLO QCD correction.

Similar study with the fiducial differential cross section work is performed for the lepton charge asymmetry. The predicted asymmetry values with total theoretical uncertainties are summarized in Table 3. To perceive the impact of NNLO corrections on the asymmetry, the difference between NLO and NNLO asymmetry predictions is investigated. Figure 6 presents the difference between NNLO and NLO asymmetry predictions for each $\eta$ bin. It is found that there is an average of 0.1% increase on NLO predictions for each PDF model.

## 3. Future Prospects on the Uncertainties

The measurement of the cross section for $W$ boson production is fundamental to the SM. High precise experimental results in $pp$ collisions have been performed using the LHC at 7 and 8 TeV center of mass energies [6–10]. Since the Large Hadron Collider (LHC) operated at $\sqrt{s}$ = 13 TeV in 2015, it is important to update 8 TeV lepton charge asymmetry and cross section results [9] with 13 TeV results. With 13 TeV LHC data, the statistical precision may be improved and the lepton charge asymmetry could be calculated for lower $x$ values. The predictions in this paper can be used to cross-check the experimental results from $pp$ collisions at 13 TeV.

In this section, the impact of the increase of the center of mass energy on theoretical uncertainties is further investigated. The changes in the theoretical uncertainties from 8 TeV to 13 TeV are as important as the change of the central values of the cross section. To investigate the impact of the center of mass energy increase on the central values and theoretical uncertainties of observable quantities, the predictions are performed with NNPDF 3.0 NNLO PDF model for $\sigma^{\text{Fid}}_{W^+}$ and $\sigma^{\text{Fid}}_{W^-}$. Table 4 shows the results for the fiducial cross section. The results show that the center of mass energy increase tends to slightly increase the uncertainties on the predictions.

## 4. Results

Observables such as the differential cross section ($d\sigma^\pm/d\eta$) and lepton asymmetry as a function of lepton pseudorapidity are shown by Figures 7 and 8. The numerical values are calculated using FEWZ 3.1 program interfaced with the three most recent PDF sets: CT14, NNPDF 3.0, and MMHT 2014. Each PDF model is plotted in different color and the colorful bands present the total uncertainties. We also cross-checked the theoretical predictions using the DYNNLO 1.5 MC tool. DYNNLO predictions are obtained using only MMHT 2014 PDF model and it is shown by black points on the upper panel of each figure.

The lower panels of Figures 7 and 8 present the differences between the central values of the different PDF

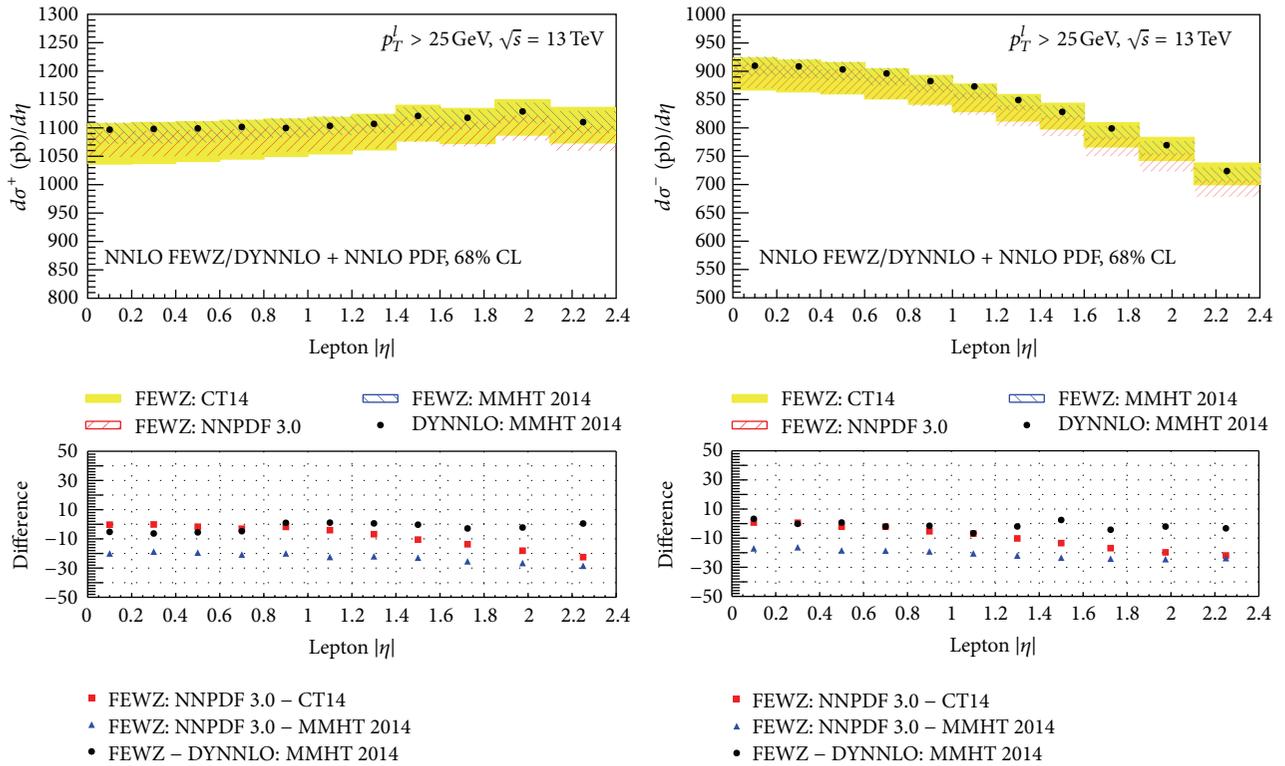

FIGURE 7: Predicted $W^{\pm}$ differential cross sections values ($d\sigma^{\pm}/d\eta$) as a function of muon pseudorapidity with NNLO QCD corrections. The predictions are calculated using the FEWZ 3.1 MC tool interfaced with different PDF sets (colorful bands) and they are compared with DYNNLO predictions (black points).

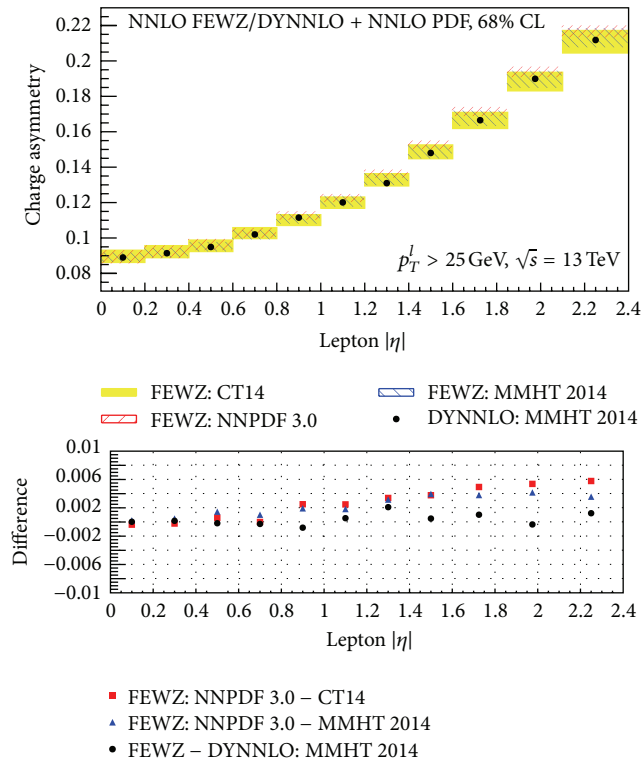

FIGURE 8: Predicted asymmetry values as a function of muon pseudorapidity with NNLO QCD corrections. The predictions are calculated using the FEWZ 3.1 MC tool interfaced with different PDF sets (colorful bands) and they are compared with DYNNLO predictions (black points).

Table 2: Predicted differential cross section numbers for CT14, NNPDF 3.0, and MMHT 2014 PDFs at $\sqrt{s} = 13$ TeV. All numbers are in units of pb.

| $|\eta|$ bin | NLO | | | NNLO | | |
|---|---|---|---|---|---|---|
| | CT14 | NNPDF 3.0 | MMHT 2014 | CT14 | NNPDF 3.0 | MMHT 2014 |
| | Values for $W^+ \to l^+\nu$ (pb) | | | | | |
| 0.00–0.20 | $1039.63^{+41.81}_{-47.83}$ | $1027.52 \pm 38.75$ | $1025.97^{+34.46}_{-35.41}$ | $1074.00^{+35.79}_{-39.71}$ | $1071.59 \pm 26.44$ | $1090.53^{+27.50}_{-25.91}$ |
| 0.20–0.40 | $1041.40^{+42.12}_{-48.10}$ | $1028.35 \pm 37.91$ | $1027.50^{+35.34}_{-36.23}$ | $1075.49^{+37.52}_{-41.51}$ | $1073.10 \pm 25.96$ | $1090.69^{+24.61}_{-22.73}$ |
| 0.40–0.60 | $1041.72^{+41.69}_{-47.70}$ | $1028.82 \pm 38.42$ | $1028.46^{+34.51}_{-35.22}$ | $1078.11^{+34.36}_{-38.58}$ | $1074.11 \pm 27.09$ | $1092.25^{+26.91}_{-24.91}$ |
| 0.60–0.80 | $1045.52^{+41.79}_{-47.81}$ | $1031.93 \pm 39.24$ | $1031.13^{+35.45}_{-35.89}$ | $1081.53^{+35.87}_{-39.92}$ | $1076.28 \pm 24.41$ | $1095.42^{+24.53}_{-22.14}$ |
| 0.80–1.00 | $1046.96^{+41.12}_{-47.00}$ | $1032.76 \pm 37.32$ | $1034.33^{+34.91}_{-35.14}$ | $1084.92^{+33.01}_{-37.22}$ | $1080.82 \pm 25.81$ | $1099.02^{+24.91}_{-22.41}$ |
| 1.00–1.20 | $1053.04^{+40.89}_{-46.55}$ | $1037.63 \pm 37.10$ | $1038.12^{+34.78}_{-34.80}$ | $1088.66^{+32.33}_{-36.54}$ | $1082.33 \pm 25.71$ | $1102.63^{+26.21}_{-23.42}$ |
| 1.20–1.40 | $1054.89^{+40.10}_{-44.91}$ | $1041.09 \pm 38.62$ | $1043.63^{+36.34}_{-36.21}$ | $1094.43^{+32.51}_{-35.89}$ | $1085.76 \pm 24.91$ | $1105.46^{+24.78}_{-21.67}$ |
| 1.40–1.60 | $1065.80^{+41.19}_{-45.15}$ | $1048.82 \pm 38.11$ | $1054.03^{+36.74}_{-36.61}$ | $1109.74^{+36.90}_{-39.15}$ | $1097.92 \pm 24.61$ | $1118.44^{+26.06}_{-22.70}$ |
| 1.60–1.85 | $1060.71^{+42.26}_{-44.94}$ | $1042.06 \pm 36.78$ | $1048.01^{+35.92}_{-36.23}$ | $1103.92^{+33.09}_{-34.22}$ | $1089.61 \pm 28.64$ | $1112.82^{+24.89}_{-21.77}$ |
| 1.85–2.10 | $1076.22^{+43.03}_{-45.31}$ | $1055.89 \pm 38.35$ | $1064.91^{+38.32}_{-38.98}$ | $1118.13^{+34.23}_{-33.94}$ | $1100.20 \pm 26.82$ | $1124.90^{+24.51}_{-25.91}$ |
| 2.10–2.40 | $1064.49^{+44.40}_{-46.42}$ | $1042.31 \pm 38.16$ | $1053.52^{+38.33}_{-39.54}$ | $1104.01^{+35.22}_{-33.85}$ | $1082.22 \pm 29.87$ | $1109.39^{+26.69}_{-24.69}$ |
| | Values for $W^- \to l^-\nu$ (pb) | | | | | |
| 0.00–0.20 | $874.12^{+35.31}_{-41.69}$ | $861.91 \pm 34.52$ | $861.29^{+31.83}_{-33.22}$ | $899.02^{+27.53}_{-34.31}$ | $896.05 \pm 26.24$ | $913.11^{+24.10}_{-24.12}$ |
| 0.20–0.40 | $869.22^{+35.18}_{-41.40}$ | $856.94 \pm 33.34$ | $856.51^{+30.89}_{-32.32}$ | $895.41^{+30.41}_{-36.67}$ | $892.24 \pm 27.32$ | $908.48^{+21.50}_{-21.59}$ |
| 0.40–0.60 | $863.83^{+34.11}_{-40.40}$ | $850.84 \pm 33.51$ | $850.20^{+29.42}_{-30.77}$ | $891.55^{+27.32}_{-34.10}$ | $885.59 \pm 26.53$ | $904.11^{+21.62}_{-21.73}$ |
| 0.60–0.80 | $854.21^{+33.62}_{-39.73}$ | $840.66 \pm 31.78$ | $841.89^{+29.56}_{-30.82}$ | $881.44^{+26.31}_{-32.78}$ | $875.56 \pm 27.91$ | $894.20^{+21.53}_{-21.50}$ |
| 0.80–1.00 | $844.38^{+33.32}_{-39.19}$ | $827.89 \pm 32.25$ | $831.91^{+29.14}_{-30.11}$ | $870.78^{+25.12}_{-31.42}$ | $862.03 \pm 26.61$ | $881.03^{+21.41}_{-21.32}$ |
| 1.00–1.20 | $827.75^{+31.43}_{-37.22}$ | $810.52 \pm 29.86$ | $817.31^{+28.11}_{-28.89}$ | $856.30^{+24.82}_{-30.62}$ | $846.14 \pm 25.12$ | $866.39^{+18.63}_{-18.32}$ |
| 1.20–1.40 | $811.44^{+30.80}_{-35.91}$ | $791.89 \pm 28.91$ | $800.14^{+28.31}_{-28.94}$ | $838.56^{+22.42}_{-27.92}$ | $825.32 \pm 22.67$ | $847.15^{+19.71}_{-19.40}$ |
| 1.40–1.60 | $795.29^{+29.80}_{-34.24}$ | $774.46 \pm 29.32$ | $784.64^{+27.51}_{-27.89}$ | $823.33^{+22.24}_{-26.92}$ | $807.43 \pm 22.91$ | $830.54^{+17.40}_{-16.90}$ |
| 1.60–1.85 | $762.13^{+28.72}_{-32.21}$ | $740.41 \pm 27.30$ | $753.51^{+26.32}_{-26.72}$ | $789.70^{+21.93}_{-25.40}$ | $770.81 \pm 20.94$ | $794.62^{+17.00}_{-17.09}$ |
| 1.85–2.10 | $736.84^{+27.89}_{-30.50}$ | $714.36 \pm 25.34$ | $727.91^{+24.72}_{-25.20}$ | $764.25^{+22.43}_{-24.74}$ | $743.11 \pm 21.89$ | $767.22^{+16.76}_{-16.25}$ |
| 2.10–2.40 | $691.84^{+25.71}_{-27.93}$ | $669.25 \pm 23.50$ | $683.12^{+23.92}_{-24.66}$ | $719.54^{+20.30}_{-21.56}$ | $696.94 \pm 19.53$ | $720.35^{+17.50}_{-17.01}$ |

Table 3: Predicted asymmetry numbers for CT14, NNPDF 3.0, and MMHT 2014 PDFs with NLO and NNLO QCD corrections at $\sqrt{s} = 13$ TeV.

| $|\eta|$ bin | $A$ (%) | | | | | |
|---|---|---|---|---|---|---|
| | NLO | | | NNLO | | |
| | CT14 | NNPDF 3.0 | MMHT 2014 | CT14 | NNPDF 3.0 | MMHT 2014 |
| 0.00–0.20 | $8.65^{+0.37}_{-0.32}$ | $8.76 \pm 0.36$ | $8.72^{+0.35}_{-0.29}$ | $8.86^{+0.47}_{-0.30}$ | $8.92 \pm 0.45$ | $8.85^{+0.39}_{-0.33}$ |
| 0.20–0.40 | $9.01^{+0.38}_{-0.33}$ | $9.09 \pm 0.29$ | $9.08^{+0.32}_{-0.25}$ | $9.13^{+0.62}_{-0.52}$ | $9.20 \pm 0.57$ | $9.11^{+0.41}_{-0.33}$ |
| 0.40–0.60 | $9.34^{+0.37}_{-0.31}$ | $9.47 \pm 0.28$ | $9.48^{+0.31}_{-0.25}$ | $9.47^{+0.53}_{-0.41}$ | $9.61 \pm 0.50$ | $9.42^{+0.40}_{-0.32}$ |
| 0.60–0.80 | $10.07^{+0.42}_{-0.36}$ | $10.22 \pm 0.25$ | $10.10^{+0.37}_{-0.31}$ | $10.19^{+0.62}_{-0.53}$ | $10.27 \pm 0.70$ | $10.11^{+0.37}_{-0.27}$ |
| 0.80–1.00 | $10.70^{+0.36}_{-0.31}$ | $11.01 \pm 0.27$ | $10.84^{+0.36}_{-0.29}$ | $10.94^{+0.40}_{-0.42}$ | $11.26 \pm 0.67$ | $11.01^{+0.42}_{-0.33}$ |
| 1.00–1.20 | $11.98^{+0.41}_{-0.37}$ | $12.29 \pm 0.21$ | $11.91^{+0.41}_{-0.35}$ | $11.94^{+0.46}_{-0.45}$ | $12.24 \pm 0.45$ | $11.99^{+0.46}_{-0.35}$ |
| 1.20–1.40 | $13.05^{+0.37}_{-0.32}$ | $13.59 \pm 0.25$ | $13.21^{+0.44}_{-0.38}$ | $13.23^{+0.51}_{-0.45}$ | $13.62 \pm 0.35$ | $13.23^{+0.45}_{-0.32}$ |
| 1.40–1.60 | $14.53^{+0.41}_{-0.35}$ | $15.05 \pm 0.26$ | $14.65^{+0.51}_{-0.45}$ | $14.81^{+0.54}_{-0.57}$ | $15.24 \pm 0.50$ | $14.77^{+0.54}_{-0.42}$ |
| 1.60–1.85 | $16.38^{+0.49}_{-0.43}$ | $16.92 \pm 0.38$ | $16.35^{+0.54}_{-0.46}$ | $16.59^{+0.51}_{-0.42}$ | $17.13 \pm 0.39$ | $16.68^{+0.54}_{-0.42}$ |
| 1.85–2.10 | $18.72^{+0.55}_{-0.50}$ | $19.29 \pm 0.29$ | $18.80^{+0.56}_{-0.45}$ | $18.79^{+0.66}_{-0.60}$ | $19.37 \pm 0.45$ | $18.90^{+0.58}_{-0.49}$ |
| 2.10–2.40 | $21.22^{+0.62}_{-0.61}$ | $21.80 \pm 0.31$ | $21.33^{+0.60}_{-0.51}$ | $21.08^{+0.64}_{-0.65}$ | $21.66 \pm 0.38$ | $21.26^{+0.61}_{-0.54}$ |

model predictions. NNPDF 3.0 is taken as a reference and the differences with CT14 and MMHT 2014 predictions are calculated for each $|\eta|$ bin. The lower panels also show the difference between FEWZ and DYNNLO predictions using MMHT 2014 PDF model. The disagreement between the FEWZ 3.1 and DYNNLO 1.5 is within 1%. Figures 7 and 8 prove that both generators have remarkable agreement with each other.

## 5. Conclusion

The LHC at CERN laboratory collided with protons at 7 and 8 TeV center of mass energies and recently it has reached 13 TeV. In this paper, QCD predictions with the most recent PDFs to date for the inclusive production of $W$ bosons in $pp$ collisions at $\sqrt{s} = 13$ TeV are presented with up to NNLO QCD corrections. The impacts of NNLO QCD corrections and NLO predictions on the central value of the observables and their theoretical uncertainties are investigated. The results show that the NNLO corrections increase the NLO CT14 and NNPDF 3.0 cross section predictions by approximately 4% while they increase MMHT 2014 cross section predictions at NLO by approximately 6%. The uncertainties associated with PDF, strong coupling constant, and high order correction scales are compared. The scale uncertainty at the NLO is reduced with NNLO QCD correction by a factor of 3.5. In addition, $\alpha_s$ uncertainty on the cross section at the NLO is slightly reduced with NNLO QCD correction.

TABLE 4: Predicted fiducial cross section numbers with NNPDF 3.0 with NNLO QCD corrections at 8 TeV and 13 TeV center of mass energies. All numbers are in units of pb. Relative uncertainties are expressed in parentheses.

|  | 8 TeV | | 13 TeV | |
|---|---|---|---|---|
|  | $W^+$ (pb) | $W^-$ (pb) | $W^+$ (pb) | $W^-$ (pb) |
| $\sigma^{\text{Fid}}$ | 4367.9 | 2916.9 | 6458.2 | 4702.5 |
| PDF uncertainty | 90.8 (2.07%) | 61.9 (2.12%) | 126.7 (1.96%) | 104.8 (2.22%) |
| Scale uncertainty | 14.2 (0.32%) | 8.0 (0.27%) | 47.7 (0.74%) | 34.0 (0.72%) |
| $\alpha_s$ uncertainty | 42.2 (0.96%) | 32.2 (1.10%) | 68.9 (1.06%) | 52.5 (1.11%) |
| Total uncertainty | 101.1 (2.31%) | 70.2 (2.40%) | 151.9 (2.35%) | 122.0 (2.59%) |

Based on the observables such as the differential cross section and lepton charge asymmetry as a function of lepton pseudorapidity, it is found that the latest PDF model predictions are in good agreement with each other unlike previous PDF sets used in the recent charge asymmetry analyses [7–10].

## Competing Interests

The authors declare that they have no competing interests.

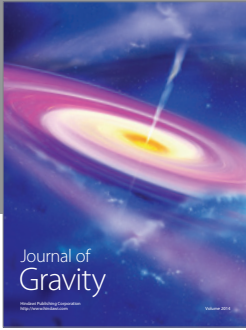

Journal of
**Gravity**

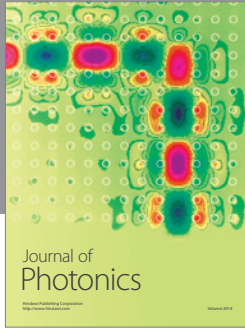

Journal of
**Photonics**

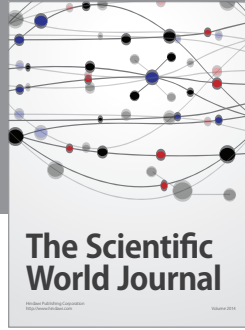

**The Scientific World Journal**

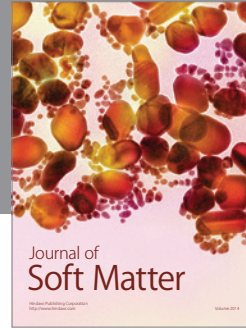

Journal of
**Soft Matter**

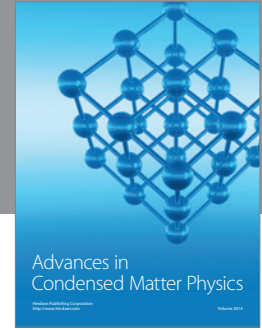

Advances in
**Condensed Matter Physics**

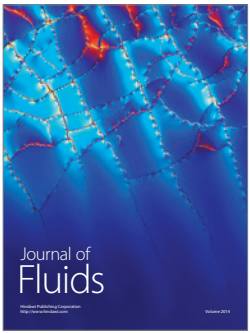

Journal of
**Fluids**

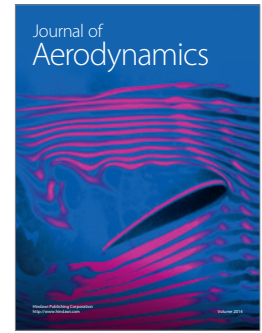

Journal of
**Aerodynamics**

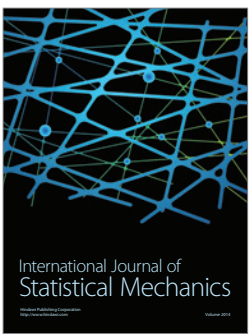

International Journal of
**Statistical Mechanics**

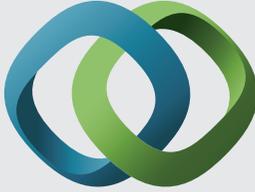

**Hindawi**

Submit your manuscripts at
http://www.hindawi.com

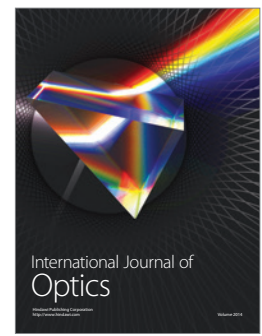

International Journal of
**Optics**

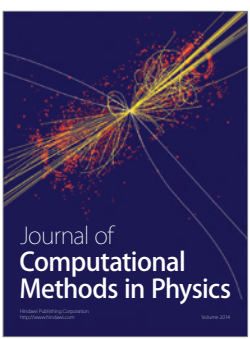

Journal of
**Computational Methods in Physics**

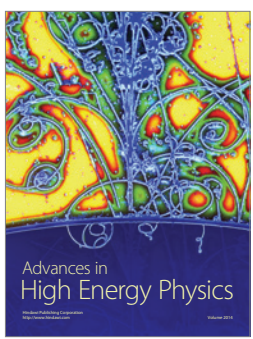

Advances in
**High Energy Physics**

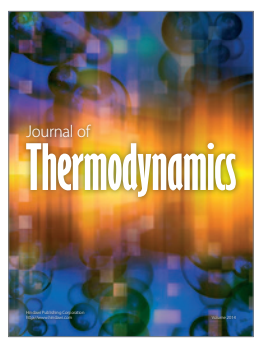

Journal of
**Thermodynamics**

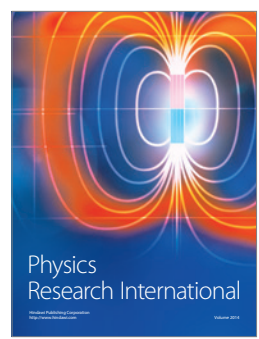

**Physics Research International**

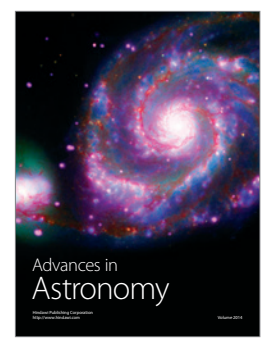

Advances in
**Astronomy**

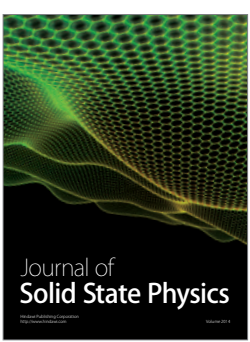

Journal of
**Solid State Physics**

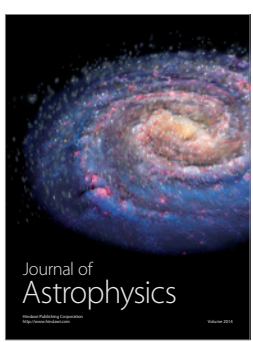

Journal of
**Astrophysics**

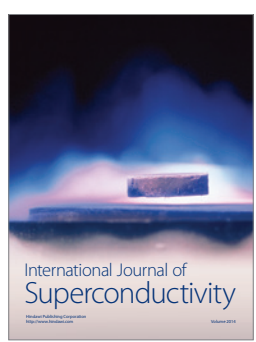

International Journal of
**Superconductivity**

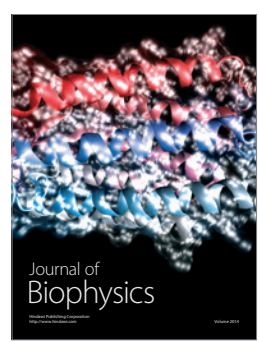

Journal of
**Biophysics**

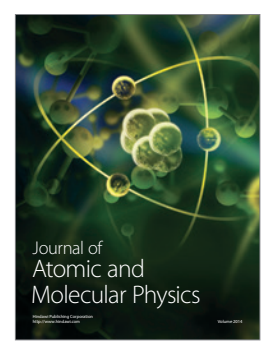

Journal of
**Atomic and Molecular Physics**